\documentclass[%
 reprint,
superscriptaddress,
 amsmath,amssymb,
 aps,
]{revtex4-2}

\usepackage{graphicx}% Include figure files
\usepackage{dcolumn}% Align table columns on decimal point
\usepackage{bm}% bold math
\usepackage{xcolor}
\begin{document}

%\preprint{APS/123-QED}

\title{Imaging anisotropic vortex dynamics in FeSe}% Force line breaks with \\

\author{Irene P. Zhang}
\affiliation{Department of Applied Physics, Stanford University, Stanford, CA 94305, USA}
\affiliation{Stanford Institute for Materials and Energy Sciences, SLAC National Accelerator Laboratory, 2575 Sand Hill Road, Menlo Park, CA 94025, USA}

\author{Johanna C. Palmstrom}
\affiliation{Department of Applied Physics, Stanford University, Stanford, CA 94305, USA}
\affiliation{Stanford Institute for Materials and Energy Sciences, SLAC National Accelerator Laboratory, 2575 Sand Hill Road, Menlo Park, CA 94025, USA}
\author{Hilary Noad}
\affiliation{Department of Applied Physics, Stanford University, Stanford, CA 94305, USA}
\affiliation{Stanford Institute for Materials and Energy Sciences, SLAC National Accelerator Laboratory, 2575 Sand Hill Road, Menlo Park, CA 94025, USA}
%\affiliation{
%\affiliation{
%\affiliation{%

\author{Logan Bishop-Van Horn}
\affiliation{Stanford Institute for Materials and Energy Sciences, SLAC National Accelerator Laboratory, 2575 Sand Hill Road, Menlo Park, CA 94025, USA}
\affiliation{Department of Physics, Stanford University, Stanford, CA 94305, USA}

\author{Yusuke Iguchi}
\affiliation{Department of Applied Physics, Stanford University, Stanford, CA 94305, USA}
\affiliation{Stanford Institute for Materials and Energy Sciences, SLAC National Accelerator Laboratory, 2575 Sand Hill Road, Menlo Park, CA 94025, USA}

\author{Zheng Cui}
\affiliation{Department of Applied Physics, Stanford University, Stanford, CA 94305, USA}
\affiliation{Stanford Institute for Materials and Energy Sciences, SLAC National Accelerator Laboratory, 2575 Sand Hill Road, Menlo Park, CA 94025, USA}

\author{John R. Kirtley}
\affiliation{Geballe Laboratory for Advanced Materials, Stanford University, Stanford, CA 94305, USA}
\author{Ian R. Fisher}
\affiliation{Department of Applied Physics, Stanford University, Stanford, CA 94305, USA}
\affiliation{Stanford Institute for Materials and Energy Sciences, SLAC National Accelerator Laboratory, 2575 Sand Hill Road, Menlo Park, CA 94025, USA}
\affiliation{Geballe Laboratory for Advanced Materials, Stanford University, Stanford, CA 94305, USA}
\author{Kathryn A. Moler}
\affiliation{Department of Applied Physics, Stanford University, Stanford, CA 94305, USA}
\affiliation{Stanford Institute for Materials and Energy Sciences, SLAC National Accelerator Laboratory, 2575 Sand Hill Road, Menlo Park, CA 94025, USA}
\affiliation{Geballe Laboratory for Advanced Materials, Stanford University, Stanford, CA 94305, USA}
\date{\today}

\begin{abstract}
Strong vortex pinning in FeSe could be useful for technological applications and could provide clues about the  coexistence of superconductivity and nematicity. To characterize the pinning of individual, isolated vortices, we simultaneously apply a local magnetic field and image the vortex motion with scanning SQUID susceptibility. We find that the pinning is highly anisotropic: the vortices move easily along directions that are parallel to the orientations of twin domain walls and pin strongly in a perpendicular direction. These results are consistent with a scenario in which the anisotropy arises from vortex pinning on domain walls and quantify the dynamics of individual vortex pinning in FeSe.
\end{abstract}

\maketitle

%\tableofcontents

%\section{\label{sec:level1}First-level heading:\protect\\ The line
%break was forced \lowercase{via} \textbackslash\textbackslash}

\section{\label{sec:level1}Introduction}
FeSe is a particularly simple layered iron based superconductor (Fe SC), with an un-strained critical temperature of about 8 K \cite{hsu2008superconductivity}, an increase in this critical temperature with pressure to 36.7 K at ~8 GPa \cite{medvedev2009electronic}, and with a report of interface-induced high-temperature superconductivity above 50 K in single unit-cell films on SrTiO$_3$ \cite{qing2012interface}. In general, iron-based superconductors exhibit complex interplay between superconducting, nematic, and magnetic orders, and undergo tetragonal-to-orthorhombic phase transitions close to magnetic ordering transitions \cite{bohmer2017nematicity}. Unlike other Fe SCs, FeSe does not magnetically order, providing an opportunity to study the superconductivity-nematicity relationship without the added complexity of an ordered magnetic state. The structural transition of FeSe occurs at 90 K and is understood to be driven by electronic nematic order \cite{hsu2008superconductivity,magadonna2008crystal,mcqueen2009tetragonal,massat2016charge}. In the orthorhombic state, FeSe and other Fe SCs form domains separated by twin boundaries (TBs). The superconducting pairing mechanism has been discussed in terms of spin-fluctuation pairing \cite{rhodes2018scaling,sprau2017discovery,wang2016strong} but the nodal character of the gap remains controversial. Some superfluid density \cite{kasahara2014field}, thermal conductivity \cite{kasahara2014field}, and tunneling spectroscopy measurements \cite{song2011direct,kasahara2014field} are consistent with line nodes in the orbital component of the superconducting order parameter; however, recent thermal conductivity \cite{bourgeois2016thermal,watashige2017quasiparticle}, STM \cite{sprau2017discovery}, and London penetration depth \cite{teknowijoyo2016enhancement} studies suggest that FeSe is fully gapped but with deep gap minima. The relationship between nematic order and superconductivity is likewise controversial, with NMR studies suggesting that nematic order competes with superconductivity \cite{baek2015orbital,baek2016nematicity} and heat capacity and thermal expansion studies suggest that it enhances superconductivity \cite{wang2017enhanced}. 
 
Although vortex pinning in superconductors is of great practical as well as fundamental interest, its mechanism is still poorly understood  \cite{blatter1994vortices}. Recently it has been possible to study this pinning directly by imaging individual vortices while manipulating them.  Auslaender {\it et al.} \cite{auslaender2009mechanics}  dragged vortices in a cuprate superconductor over distances of a few microns using a magnetic force microscope tip. They found an enhanced response of the vortex to pulling when the tip was oscillated transversely. They also found enhanced vortex pinning anisotropy, which they attributed to clustering of oxygen vacancies in their sample. Later work from Shapira {\it et al.} used MFM to drag vortices along twin boundaries in  YBa\textsubscript{2}Cu\textsubscript{3}O\textsubscript{t-$\delta$} and demonstrated that the vortices moved in a series jumps, consistent with power-law behavior \cite{shapira2015disorder}. Kalisky {\it et al.} \cite{kalisky2011behavior} showed that vortices, when dragged by a scanning SQUID microscope, avoided crossing twin boundaries in underdoped Ba(Fe$_{1-x}$Co$_x$)$_2$As$_2$. Embon {\it et al.} \cite{embon2015probing} used a SQUID on a tip to image the movement over a few tens of nanometers of vortices driven by  applied supercurrents in a thin Pb film. They were able to map out anisotropic and spatially inhomogeneous pinning forces on the vortices, which they attributed to multiple overlapping pinning sites. 
 
The vortex pinning properties of FeSe have attracted interest \cite{sun2015critical, sun2015enhancement, leo2015vortex, massee2015imaging} and make it a potential competitor to high-$T_c$ cuprates for high field applications \cite{leo2015vortex}. Critical current density studies of vortex pinning in FeSe have found that it is dominated by strong point-like pinning \cite{sun2015critical}, while STM studies have shown that vortices preferentially pin on TBs in FeSe, where the superfluid density is reduced \cite{song2012suppression}. In this paper we present scanning SQUID magnetometry and susceptibility images of vortices trapped in single crystals of FeSe. The susceptibility images show structures that we attribute to motion of the pinned vortices driven by the magnetic fields applied by the field coil integrated into our SQUID susceptometers. Analysis of our data using a simple model is consistent with a quadratic dependence of the restoring force on displacement along the direction of the TBs, with a much larger restoring force in the orthogonal direction. The anisotropy in the in-plane restoring forces can be as large as a factor of 20.

\section{Materials and Methods}

\begin{figure}
\centering
\includegraphics[width=\linewidth]{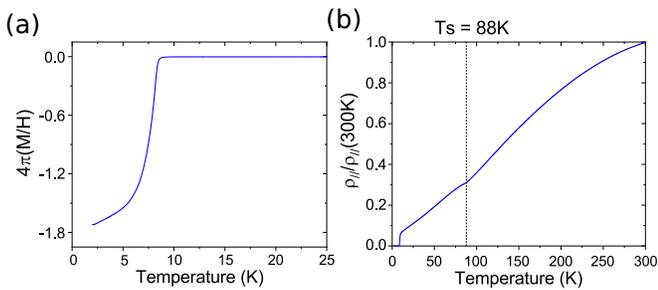}
\caption{Magnetic-susceptibility (a) and resistivity (b) curves used to determine $T_c$ = 8.8 K and $T_s$ = 88 K, respectively. In (b), the in-plane resistivity $\rho_{||}$ is plotted, normalized by its value at room temperature.} 
\label{fig:TsTccurves}
\end{figure}

We used scanning SQUID microscopy to image magnetometry and susceptibility 
in bulk single crystal FeSe. Single crystal FeSe samples were grown by 
chemical vapor transport following the growth procedure outlined in Ref. \onlinecite{boehmer2016growth} and exfoliated with Kapton tape and silver paint (Dupont 4929N) as in Ref. \onlinecite{massat2017} to achieve a surface flat enough 
to be scanned using our susceptometers. The superconducting transition and structural transition temperatures of the batch of samples used were found to be 8.2 K and 88 K respectively. The bulk superconducting transition temperature was extracted from magnetic-susceptibility measurements taken using the vibrating sample mount option of the MPMS 3 from Quantum Design. The structural transition temperature was determined from resistivity measurements on a free standing crystal taken using the MPMS 3 paired with a Linear Research Model LR-700 AC resistance bridge. The susceptibility and resistivity data are shown in Fig. \ref{fig:TsTccurves}.

Our SQUID susceptometers contain two Nb pickup loop/field coil pairs 
arranged in a gradiometric layout \cite{KirtleyRevSciInstrum16}. The pickup loop and 
field coil are covered by Nb shielding so that flux passes only through the 
loop and not through the gaps between the leads. The inner radius of the 
pickup loop was 0.3 um, resulting in sub-micron spatial resolution. As the 
susceptometer scans across the surface of the sample, we record the magnetic
flux passing through the pickup loop. The dc signal is recorded as 
magnetometry and is reported in units of the flux quantum $\Phi_0 = h/2e$. 
We use an SR830 lock-in amplifier to pass an ac current through the field 
coil, creating a local magnetic field, and record the ac flux at that 
frequency to measure susceptibility. The gradiometric design cancels out the
flux due to the field coil so that the pickup loop only measures the 
magnetic response of the sample. The susceptibility is normalized by the 
lock-in amplifier current and is reported in units of $\Phi_0$/A. This 
design allows us to image local magnetic fields and susceptibility at the 
surface of a sample simultaneously.

\section{Results}
\subsection{Imaging vortex motion}
\begin{figure}
\centering
\includegraphics[width=\linewidth]{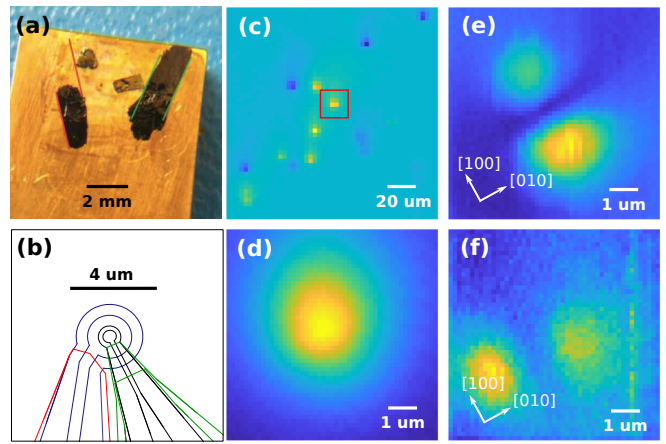}
\caption{Images of vortices and vortex motion in FeSe. (a) Optical microscope image of the two samples imaged for this paper. The black mark with red lines indicates the approximate tetragonal [100] directions for Sample 1 (right) and the black mark with green lines indicate the same for Sample 2 (left). (b) Layout of the pickup-loop/field coil geometry for the SQUID susceptometer used. (c) Large area magnetometry image of the surface of FeSe. The square outlines the area imaged in (d) and (e). (d) Magnetometry image of a single vortex. The full-scale variation of the false color look-up table corresponds to 31m$\Phi_0$ magnetic flux through the SQUID pickup loop. (e) Susceptibility image taken simultaneously with the magnetometry image in, showing a butterfly in one of two types of domains (d). The full-scale variation here is 4.3$\Phi_0/A$. The white arrows indicate the FeSe tetragonal $a$ crystal axes directions. (f) Susceptibility image of a second vortex in the second TB direction in the same sample as (c-e). Full-scale variation 0.57$\Phi_0/A$.} 
\label{fig:vortex_images}
\end{figure}
In four measured samples, we found two-lobed features ("butterflies") in the susceptibility 
images, accompanying superconducting vortices. Not all vortices had these 
``butterflies" and their brightness varied from vortex to 
vortex. Increasing the sample temperature increases the brightness of the 
lobes up to Tc, after which both the vortex and accompanying butterflies 
disappear. The butterflies orient along one of two directions which are 
perpendicular to each other. Further, the relative brightness of the two 
lobes is consistent within butterflies of the same orientation, and the 
dimmer lobes are located in areas opposite to the location of the field coil
shielding. The brightness of the butterflies was varied from vortex to 
vortex, ranging from around one tenth of a $\Phi_0/A$ to a few $\Phi_0/A$. 
Besides wiggling the vortex, the field coil could also push the vortex to a different pinning site. Not all vortices had butterflies, likely because they
were pinned too strongly to show a signal in susceptibility. Due to
unevennesss in the sample surface, it was difficult to control for the
height and angle of the pickup loop/field coil pair relative to the surface
area being scanned.
We used x-ray diffraction to determine the tetragonal crystal 
axes in Samples 1 and 2 and found that a line cutting through the two lobes orients the butterflies either along or perpedicular to the tetragonal [100] direction. This orientation is 
consistent with the butterflies being aligned with TBs. Fig. \ref{fig:vortex_images} shows two representative susceptibility butterflies from Sample
1, along with the corresponding magnetometry image showing the vortex for 
the first butterfly. A background is subtracted from all susceptibility 
images by fitting a plane to the four corners of the scan. 

\begin{figure}
\centering
\includegraphics[width=\linewidth]{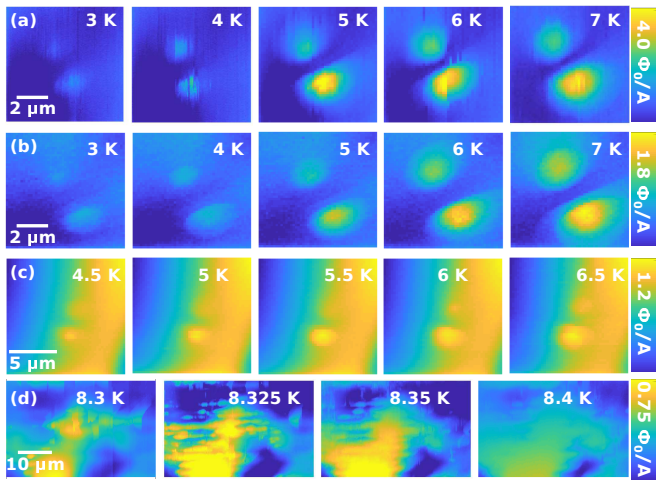}
\caption{Temperature series of susceptometry images for two butterflies in Sample 1 (a)-(b) and one in Sample 2 (c), and temperature series in Sample 3 showing striped features in susceptometry along the TB direction (d).} 
\label{fig:Tseries}
\end{figure}

In Fig. \ref{fig:Tseries}, we show a series of susceptibility scans for three vortices with butterflies at variouss temperatures, as well as a temperature series showing the development of stripes in diamagnetic susceptibility close to $T_c$. Near the superconducting transition, the vortex becomes easier to move and the features of the butterflies become sharper. In one sample we also observed striped variations in susceptibility run along the tetragonal [100] direction.

\subsection{Modeling vortex motion}

We model our susceptibility images by 1) calculating the magnetic fields inside the superconductor at the vortex position due to the applied currents through the field coil, 2) calculating the motion of the vortex in response to these fields using a simple model with an anisotropic pinning potential, and then 3) calculating the change in flux through the susceptometer due to the vortex motion. The SQUID susceptibility is given by the response flux $\Phi$ divided by the field coil current $I$.
\subsubsection{Applied fields}

Consider a geometry in which a scanning SQUID susceptometer, composed of niobium films with penetration depth $\lambda_{\rm Nb}$, with layout in the pickup loop/field coil region as illustrated if Fig. \ref{fig:vortex_images}b, is assumed oriented parallel to the sample surface, has a spacing $z_0$ between the surface of the susceptometer and the surface of the sample, and is in the half-space $z>0$ (region 1). The superconducting sample, with penetration depth $\lambda$, is in the half-space $z<0$ (region 2). The fields generated by the susceptometer are calculated following Ref. \onlinecite{brandt2005thin} as described for our sensors in Ref. \onlinecite{KirtleyRevSciInstrum16}. For a thin superconducting film with thickness $t < 
\lambda_{Nb}$, one can define a stream function $g$ by ${\bf J} = {\bf \hat{z}} 
\times \nabla g$, where ${\bf J}$ is the sheet current \cite{brandt2005thin}. One interpretation of the stream function is that it defines a density of magnetization in the $\hat{z}$ direction, or equivalently a collection of small current loops in the plane. Once the stream functions $g_{j,l}$ are known for all of the grid points $j$ in all of the superconducting layers $l$ in the susceptometer, the source potential at any point $i$ in the half-space $z>0$ but outside of the susceptometer superconducting layers is given by \cite{kogan2003meissner}
\begin{equation}
\varphi_s(\vec{r}_i,z_i)=-\sum_l\sum_j \frac{w g_{j,l}}{4\pi}\frac{z_i-z_{j,l}}{((z_i-z_{j,l})^2+\rho_{i,j,l}^2)^{3/2}},
\label{eq:source}
\end{equation}
where $w$ is the area of the pixels used, and $\rho_{i,j,l}=\sqrt{(x_i-x_{j,l})^2+(y_i-y_{j,l})^2}$. For what follows we calculate the source potential $\varphi_s$ for all $\vec{r}_i$ and $z=0$.

Once the source fields are known, we match boundary conditions at $z=0$. This is done by expanding the scalar magnetic potential $\varphi_1(\vec{r},z)$ outside the sample ($z>0$) and the magnetic field $\vec{H_2}(\vec{r},z)$ inside the sample ($z<0$) in Fourier series that are constructed to satisfy Maxwell's equations for $z>0$ and London's equation for $z<0$ \cite{kogan2003meissner}.

\begin{equation}
\varphi_1(\vec{r},z)= \frac{1}{(2\pi)^2} \int d^2\kappa(\varphi_s(\vec{\kappa})e^{\kappa z}+\varphi_r(\vec{\kappa})e^{-\kappa z})e^{i \vec{\kappa}\cdot\vec{r}}  
\label{eq:phi}
\end{equation}

\begin{equation}
\vec{H}_2(\vec{r},z) =\frac{1}{(2\pi)^2} \int d^2 \kappa \, \vec{h}_2(\vec{\kappa}) e^{q z} e^{i \vec{\kappa} \cdot \vec{r}},
\label{eq:H}
\end{equation}
where $\vec{\kappa}=\kappa_x \hat{x}+\kappa_y \hat{y}$, $\kappa = \sqrt{\kappa_x^2+\kappa_y^2}$, $q^2=\kappa^2+1/\lambda^2$ and $\vec{H}_1 = \vec{\nabla}\varphi_1$. Using the boundary conditions $\vec{B}\cdot \hat{z}$ and $\vec{H} \times \hat{z}$ continuous at $z=0$, as well as $\vec{\nabla}\cdot \vec{B_2}=0$ and $\vec{B}=\mu_0 \vec{H}$ results in \cite{kogan2003meissner}

\begin{equation}
h_{2,z}(\vec{\kappa}) = \frac{2\kappa^2}{\kappa+q} \varphi_s(\vec{\kappa})
\label{eq:h2kz}
\end{equation}

\begin{equation}
\vec{h}_{2,||}(\vec{\kappa})=\frac{2 i \vec{\kappa} q}{\kappa+q}\varphi_s(\vec{\kappa})
\label{eq:h2kpar}
\end{equation}
The magnetic field $\vec{H}_2(\vec{r},z)$ penetrating the superconductor is calculated taking the Fourier transform of $\vec{h}_2(\vec{\kappa})$ (Eq. \ref{eq:H}). 

\subsubsection{Vortex motion}
The additional magnetic energy $\delta u$ of the vortex per unit length due to the susceptometer applied field is given by
\begin{equation}
    \delta u(\vec{r}) = \frac{1}{2} \int d^2r' \, \vec{H}_{2}(\vec{r}') \cdot \vec{B}_V(\vec{r}-\vec{r}')
    \label{eq:du}
\end{equation}
where $\vec{B}_V(\vec{r}'-\vec{r})$ is the field of the vortex centered at $x,y$. If we assume the vortex is oriented in the $\hat{z}$ direction and neglect the finite extent of the vortex fields, this becomes simply
\begin{equation}
    \delta u(\vec{r}) = \Phi_0 H_{2,z}(\vec{r})/2
    \label{eq:du2}
\end{equation}
The total additional energy $\delta U$ of the vortex due to the field penetration into the superconductor is then
\begin{equation}
    \delta U(x,y) = \frac{\Phi_0}{2(2\pi)^2} \int_0^{-\infty} dz \, \int d^2\kappa \, h_{2,z}(\vec{\kappa}) e^{q z} e^{i \vec{\kappa} \cdot \vec{r}}
    \label{eq:dUfinal}
\end{equation}
The integration over $z$ can be done analytically, resulting in
\begin{equation}
    \delta U(x,y) = \frac{\Phi_0}{2(2\pi)^2} \int d^2\kappa \, h_{2,z}(\vec{\kappa}) e^{i \vec{\kappa} \cdot \vec{r}}/q
    \label{eq:dUmorefinal}
\end{equation}
The force on the vortex due to the susceptometer fields is the gradient of this extra potential:
\begin{equation}
    \vec{F}_{\rm SQUID}(x,y) = \vec{\nabla}(\delta U(x,y))
    \label{eq:force}
\end{equation}

We find that the force is reduced when the field coil shield is above all
or part of the vortex, meaning that when comparing scans to the 
corresponding susceptometer layout, the areas opposite to the field coil 
shield will show reduced signal.

Because we do not know the exact form or mechanism of the vortex pinning potential in FeSe, we 
use a simple quadratic model with spring constants $k_w$ and $k_s$  ($k_s \geq k_w$) associated with orthogonal axes $\hat{w}$ and $\hat{s}$ rotated by an angle $\theta$ in the $ab$ plane relative to the scan axes $\hat{x}$ and $\hat{y}$ . When the susceptometer scans in the $xy$ plane relative to the vortex position, the susceptometer applied fields pull the vortex towards (or away) from the pickup loop/field coil. Since we modulate the current through the field coil at about 1 kHz, we assume that the vortex response is much faster than the applied force, and therefore that the displacement of the vortex ${\vec{dr}} = dw\, {\hat{w}}+ds\,{\hat{s}}$ from its equilibrium position can be calculated from the balances of forces condition
\begin{equation}
    \vec{F}_{\rm SQUID} = k_s ds \, \hat{s} + k_w dw \,\hat{w}.
\end{equation}
\subsubsection{Response flux due to vortex motion}

The expected ac flux from the vortex motion is calculated using the 
gradient of the dc flux through the susceptometer pickup loop due to the vortex:
\begin{equation}
    \Phi_{ac} = \frac{d\Phi_{dc}}{dx} dx + \frac{d\Phi_{dc}}{dy} dy
\label{eq:phiac}    
\end{equation}
The flux due to a vortex is calculated following the procedure in Ref. \onlinecite{KirtleyRevSciInstrum16}; %\textcolor{green}{(From Irene: Not sure who asked this, but the model takes only the axis of the butterflies and calculates everything else from Kogan+Kirtley and Brandt)}  
the derivatives in Eq. \ref{eq:phiac} are taken numerically, and the susceptibility is calculated by dividing $\Phi_{ac}$ by the applied field coil current.
\begin{figure}
\centering
\includegraphics[width=\linewidth]{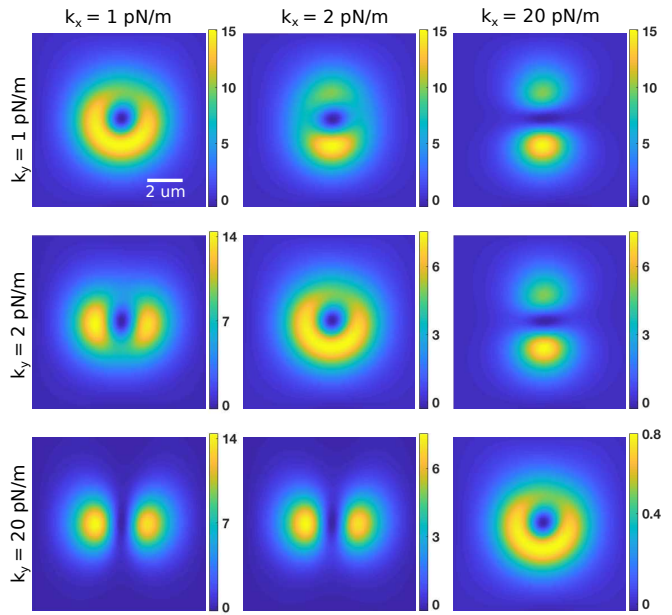}
\caption{Results of our model of vortex susceptibility images, using the susceptometer layout of Fig. \ref{fig:vortex_images}b with $z_0=2\mu m$. Unlike in the SSM data, the axes of anisotropy are chosen to lie along the image axes. The spring constants along the x- and y-axes are varied from 1 $\times10^{-9}$ N/m to 20 $\times10^{-9}$ N/m. In the cases where there is anisotropy the weak axis $k$ sets the scale of the signal while the strong axis $k$ changes the shape of the lobes. The color-scales are in units of $\Phi_0/A$. }
\label{fig:butterfly_modeling}
\end{figure}

\subsubsection{Results of the model}

Example predictions for the ac flux due to vortex motion with the susceptometer geometry 
in Fig. \ref{fig:vortex_images} is shown in Fig. \ref{fig:butterfly_modeling} for various assumed values of $k$ along the x- and y-axes of the images, with $z_0$=2 $\mu$m. The resulting 
shapes for the isotropic cases ($k_s = k_w$) are similar to an incomplete 
torus, while the shapes for the anisotropic cases ($k_s \neq k_w$) are 
distinctly lobed. The weaker axis ($k_w$) spring constant determines the intensity of the susceptibility signal,
while the stronger axis spring constant ($k_s$) largely determines the shape. As the 
strength of the strong axis spring constant is increased, the lobes become more axially symmetric, and the dark region in the center fades. Consistent with the force
profile, the lobes located across from the field coil shielding are dimmer 
than their partners. Crucially, we find that the apparent axis of the 
anisotropy is not qualitatively changed by the asymmetric susceptometer geometry.

\begin{figure}
\centering
\includegraphics[width=\linewidth]{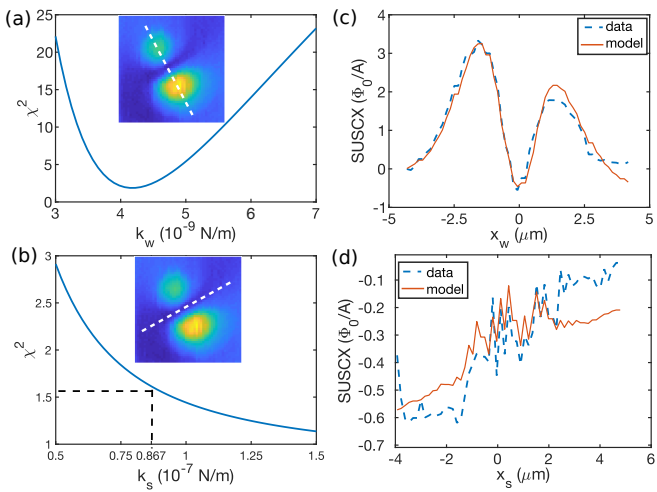}
\caption{Fits of the model to experiment using the spring constants $k_s$ and $k_w$ as fitting parameters. (a) The dependence of the $\chi^2$ difference between model and experiment along the weak axis (dashed line in inset) on $k_w$ using a fixed value of $k_s$=1 N/m. The $\hat s$ axis is assumed to be rotated by 28 degrees relative to the scanned $x$ axis, with $z_0$= 2 $\mu$m. (b) The dependence of $\chi^2$ on $k_s$ computed for cross sections through the strong axis (dashed line in inset) for a fixed value of $k_w$= $4.1\times10^{-9}$ N/m.  The black dashed line indicates the value of $k_s$ at which $\chi_2$ is doubled from its minimum value. (c) - (d)  Experimental (blue dashed line) and best-fit model (red solid line) cross-section along the weak (c) and strong (d) axes.}
\label{fig:cross_sections}
\end{figure}

\begin{figure}
\centering
\includegraphics[width=\linewidth]{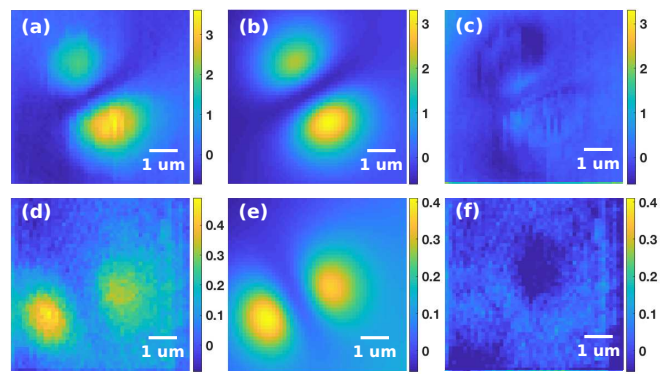}
\caption{Susceptibility data (a), best fit model (b), and difference (c) for the butterfly shown in \ref{fig:vortex_images}e. Data (d), model (e), and difference (f) for the butterfly shown in \ref{fig:vortex_images}f. The color maps are in units of $\Phi_0/A$.}
\label{fig:2d_fits}
\end{figure}

Using reduced chi-square fitting, we calculated the optimal spring constant 
in the weak axis (“$k_w$”) direction and a lower bound for the spring 
constant in the strong axis direction (“$k_s$”) for 12 butterflies from two samples. We find 
that in all cases the signal along the hard axis is so weak compared to the noise level 
that the optimal $k_s$ value approaches infinity. In Fig. \ref{fig:cross_sections} we show the fitting process for the butterflies shown in Fig. \ref{fig:vortex_images}. We fit $k_w$ and $k_s$ separately by taking cuts along the weak and strong axes, respectively. Full images for the data, model, and difference between data and model are shown in Fig. \ref{fig:2d_fits} The difference in brightness between model and experiment in the dimmer lobe can be explained by the limited spatial extent of the field coil shielding in the model SQUID geometry.

Taking the optimal $k_w$
value and the lower limit for $k_s$ results in a lower limit for the 
ratio $k_s/k_w$, which we use to characterize the anisotropy of vortex motion. The range of $k_s/k_w$ varies from around 4 to over 20, and three vortices show ratios above 20. The results are summarized in Table \ref{tab:ktable}, along with the temperature at which the data were taken.

\begin{table}[b]
\caption{\label{tab:ktable}
Summary of measured $k_w$ and $k_s$ values.}
\begin{ruledtabular}
\begin{tabular*}{\textwidth}{c @{\extracolsep{\fill}} cccc}
\multicolumn{5}{ c }{\textrm{Sample 1}}\\
\textrm{Axis}&
\textrm{$k_w$ (N/m)}&
\textrm{$k_s$ (N/m)}&
\textrm{$k_s/k_w$}&
\textrm{T (K)}
\\
\colrule
1 & $4.1\pm0.5\times10^{-9}$ & $>8.5\times10^{-8}$ & $>21$ & 7\\
1 & $8.3\pm0.9\times10^{-9}$ & $>1.7\times10^{-7}$ & $>22$ & 7\\
1 & $2.7\pm0.8\times10^{-8}$ & $>1.2\times10^{-7}$ & $>4.5$ & 7\\
2 & $4.0\pm0.9\times10^{-8}$ & $>6.2\times10^{-7}$ & $>16$ & 7.5\\
2 & $8.6\pm3.3\times10^{-8}$ & $>3.6\times10^{-7}$ & $>4.2$ & 6\\
2 & $3.2\pm0.7\times10^{-8}$ & $>1.5\times10^{-7}$ & $>4.7$ & 7\\
\hline
\multicolumn{5}{ c }{\textrm{Sample 2}}\\
\textrm{Axis}&
\textrm{$k_w$ (N/m)}&
\textrm{$k_s$ (N/m)}&
\textrm{$k_s/k_w$}&
\textrm{T (K)}
\\
\colrule
1 & $8.5\pm1.9\times10^{-9}$ & $>6.6\times10^{-8}$ & $>7.8$ & 6\\
1 & $1.7\pm0.3\times10^{-8}$ & $>6.4\times10^{-8}$ & $>3.8$ & 6\\
1 & $6.6\pm1.4\times10^{-9}$ & $>8.6\times10^{-8}$ & $>13$ & 6\\
1 & $8.0\pm2.0\times10^{-9}$ & $>3.6\times10^{-8}$ & $>4.5$ & 4.5\\
1 & $9.5\pm2.4\times10^{-9}$ & $>2.0\times10^{-7}$ & $>21$ & 7\\
2 & $1.2\pm0.4\times10^{-8}$ & $>7.4\times10^{-8}$ & $>6.2$ & 4\\
\end{tabular*}
\end{ruledtabular}
\end{table}

\begin{figure}
\centering
\includegraphics[width=\linewidth]{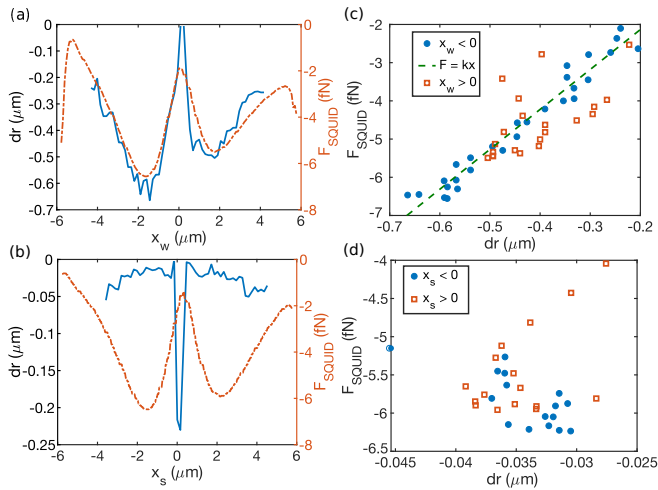}
\caption{(a)-(b) Cross sections of vortex displacement $dr$ (blue solid line) extracted from susceptibility data and simulated susceptometer force (red dashed line), for a 1 mA field coil current, along the weak (a) and strong (b) axes of the butterfly in \ref{fig:2d_fits}. The displacement tracks with the strength of the force from the susceptometer along the weak axis. For the strong axis, a linear background was subtracted from the susceptbility cross section before calculating vortex displacements. A small non-zero background leads to a finite $dr$ which does not track with $F_{\rm SQUID}$. (c) Susceptometer force versus vortex displacement along the weak axis plotted separately for the brighter (blue circles) and dimmer (red squares) lobes. A linear model (green dashed line) was used to fit the data to calculate an effective $k_w$. (d) Susceptometer force plotted against vortex displacement for the strong axis cross section. We do not see any apparent correlation between applied force and calculated displacement.}
\label{fig:forces}
\end{figure}

Another way to infer the vortex dynamics in these samples is to directly 
extract the vortex displacement from scans. This approach avoids the 
assumptions of a toy model potential entirely. Providing that the direction 
of the vortex motion is known, we can use Equation \ref{eq:phiac} and the gradient from the magnetometry image to obtain the maximum vortex displacement. In this analysis we 
exclude the values close to the vortex center and far from the vortex, as
the magnetometry gradient is very small in these areas and amplifies any 
noise in the susceptibility. This is the case along the lines where $F_s = 
0$ and $F_w = 0$, respectively. Furthermore, since the motion in the $\hat s$
direction is severely limited, we can assume the vortex moves mostly in the 
$\hat w$ direction. The motion along lines through $F_s = 0$ and with constant but 
small $F_w$ is therefore mostly radial. In Fig. \ref{fig:forces} we show line cuts of the 
experimental vortex displacement and calculated susceptometer force along the weak and strong axes of 
a butterfly from Sample 1. Because the signal along the strong axis is 
dominated by the background, we subtracted a linear fit from susceptibility 
line cuts along this axis. The absolute magnitude of the vortex displacement is at least an order of 
magnitude smaller in the strong-direction compared to the weak-direction. 
Furthermore, while the displacement along the weak axis tracks with 
susceptometer force, the displacement along the strong axis appears to be 
largely independent of the force applied by the field coil. A linear fit to 
the weak axis data is plotted in Fig 1(b). The effective $k_w$ in this case 
is $1.0 \pm 0.06 \times 10^{-8}$ N/m, about 2.5 times the optimal $k_w$ found by
fitting the data to the quadratic model. These results 
are consistent with a much larger spring constant along the strong axis than along
the weak axis, and also consistent with a linear 
force-displacement relation in the weak direction (Fig. \ref{fig:forces}(b)).

\section{Discussion}

The ac current through the field coil generates an oscillating local magnetic field and superconducting screening currents
that pull and push the vortex towards 
and away from the center of the pickup loop/field coil.  
%Ignoring time scale effects, the vortex is closest to the center of the 
%pickup loop/field coil at the field coil current maxima. 
As a result, as long as the vortex response to the applied force is faster than the time variation of the field coil current, the flux through the pickup loop is in-phase with 
the field coil current, as
reflected in the real part of the susceptibility image. This scenario is rather general to 
scanning SQUID measurements with simultaneous magnetometry and 
susceptibility imaging and does not rely on the particulars of the FeSe 
samples, except that in FeSe the motion is both large and highly 
anisotropic.

We can first rule out that the motion is isotropic but that the observed lobed shape is 
simply due to the SQUID susceptometer geometry based on the simulations in Fig 
2. The expected ac signal in the isotropic case is proportional to the 
susceptometer force, which is mostly radially symmetric except for shielded 
regions. The force is small when the pickup loop/field coil center is above 
the vortex, increases to a maximum at a distance equal to the field coil 
radius and then decreases again as the field coil moves away from the 
vortex. This force profile results in the ``doughnut" shape discussed 
earlier. The effect of increasing the spring potential along one axis is to 
weaken the effect of the force along that axis so that as k becomes very 
large the signal along that axis becomes indiscernible, producing lobed 
features. 

As noted in the introduction, previous MFM and scanning SQUID microscopy measurements have measured vortex
dynamics in superconductors\cite{auslaender2009mechanics,kalisky2011behavior,embon2015probing}. 
%In particular, an SSM study of 
%Ba(Fe\textsubscript{1−x}Cox)\textsubscript{2}As\textsubscript{2}, a sister 
%compound of FeSe, found that vortices in this compound move preferentially 
%when manipulated by the SQUID susceptometer, being dragged along twin domain
%boundaries without crossing them []. An MFM study similarly showed that 
%vortices in YBa\textsubscript{2}Cu\textsubscript{3}O\textsubscript{6.991} 
%could be dragged more easily in one direction than in the other and linked 
%this anisotropy in pinning to oxygen vacancies in the sample []. 
In contrast to the Auslaender {\it et al.} and Kalisky {\it et al.} studies, in
which the vortices were dragged to new locations on the sample, in our case and for small displacements in the Embon {\it et al.} \cite{embon2015probing} study the vortex returns to its original location and is instead oscillated in place. In the Embon {\it et al.} study the displacements are typically a few tens of nanometers, the forces are pico-Newtons, and the spring constants are $\sim 10^{-5}$ N/m, whereas in the present study the displacements are typically one micron, the forces are typically femto-newtons, and the weak spring constants are typically $10^{-8}$ N/m, 3 orders of magnitude weaker in FeSe than in Pb.

A 2012 STM study \cite{song2012suppression} found that in thin film FeSe, superconductivity is suppressed along twin domain boundaries and that vortices tend to pin on these boundaries. The cause of this suppression is thought to be due to the increased height of the Se atoms along the boundary. We did not observe chains of vortices along any direction; however, it is possible our vortex density was not high enough to observe such patterns. In three samples for which we were able to determine the direction of the crystalline axes, the vortex motion was highly anisotropic along the TB directions, suggesting that it is easier to pull vortices along the TBs than acorss them. Further evidence of this can be found in the striped variations in susceptibility signal close to $T_c$ which run along the same direction as the weak axis of the butterflies. Given that the only other known symmetry-breaking through lattice orthorhombicity, for which the crystal axes are oriented at 45 degrees to the butterfly and stripe direction, the only reasonable conclusion is that the vortices are pinned on TBs and the variations susceptibility reflect the suppressed superconductivity on the boundaries.

\section{Conclusion}

In conclusion, scanning SQUID susceptibility can be used to image the motion of vortices under the influence of locally applied magnetic fields. We have applied this method in the instance of vortices in FeSe. Detailed calculations of the magnetic fields generated by the susceptometer, combined with a simple model for the pinning forces on the vortex, show that these pinning forces can be highly anisotropic and is consistent with a quadratic dependence of the restoring force on displacement. We calculate an effective spring constant for the weak and strong axes and show that this is consistent with vortex pinning that is strong across TBs and weaker along them.

\begin{acknowledgments}
We thank J. Straquadine and E. Rosenberg, and C. Watson for experimental assistance and P. Massat for feedback on the manuscript. This work was supported by the Department of Energy, Office of Science, Basic Energy Sciences, Materials Sciences and Engineering Division, under Contract No. DE-AC02-76SF00515
\end{acknowledgments}

\bibliographystyle{apsrev4-2}
\bibliography{references}
\end{document}